# X-ray diffraction strains in laser-ablated aluminum, nickel, sodium and Invar: pressures to 475 GPa

by

S. J. Burns[1,2,†] and Danae N. Polsin[2,3]

Affiliations:

[1]Materials Science Program, The College of Arts & Science and The College of Engineering, University of Rochester, Rochester, New York, 14627 USA
[2]Department of Mechanical Engineering, 236 Hopeman Hall, University of Rochester, Rochester, New York, 14627 USA
[3]Laboratory for Laser Energetics, University of Rochester, Rochester, New York, 14623 USA
[†]Corresponding author; email: stephen.j.burns@rochester.edu

Date: August 28, 2025

## *Abstract*

Dynamically compressed materials in longitudinal waves are described by two physical models: hydrostatic pressure, with equal, normal, principal stresses or material uniaxially strained in the wave propagation direction. These models are disparate, so experimental comparisons and evaluations are important. Polycrystalline material in a state of hydrostatic pressure, will have no eccentricity of X-ray diffracted Debye-Scherrer rings. A general three-dimensional solution of Bragg diffracted X-rays based on principal crystallographic strains in the compression wave was found. The distortion of X-ray diffraction beams has been used for strain measurements; the analysis developed incorporates a strained reciprocal lattice and the incident X-ray beam. Strain distorted Polanyi surfaces form an annulus of compression with an ellipsoid of revolution in reciprocal space which is intersected by Ewald's sphere for Bragg diffraction. The *in-situ* measurements for strain describe nanosecond diffraction evaluated using two planes, $(hkl)$ and $(h'k'l')$ both in the same crystallographic phase. Diffraction from Al, Ni, Na, and Invar quantify the compression axial strains in these materials: the compression axial ratios are 0.65, 1.05, 0.88 and 1.58 at pressures of 291, 402, 409, and 367 GPa for the respective materials. Crystal structure transformations with homogeneous pressurized stresses, mandating equal normal strains, should not be anticipated to agree with heterogeneous, uniaxially strained and sheared crystalline phases. Measurements support in-plane strains increasing with pressure, $p$, in fcc and hcp aluminum as $\varepsilon_{22} = -0.24 x 10^{-3} p$ with $p$ in GPa.

## Introduction

A major paradox in laser-ablated compression of solids is finding the compatibility and consistency of two models often used to describe the compression: the first model hypothesizes that the mechanical stresses in the solid are hydrostatic i.e., a pressure [1-6]. The second model which is widely used in wave propagation is uniaxial strain in the direction of the propagating wave [7-10]. These two models are incompatible. The analysis presented based on strain will simultaneously combine measurable characteristics from both models and aid in our understanding of material behavior under extreme compression. Strain differences based on evidence from two diffraction measurements determine principal strains. If uniform strains are established in two perpendicular directions for laser-compression, the measurements suggest induced hydrostatic strains. Both models require two or more simultaneous independent strain measurements and an assumption of isotropy to find the strain tensor. Three independent strain measurements on a flat surface are needed to determine the principal strains and orientation for a 2-dimensional strain tensor [11].

X-rays are frequently used to measure interatomic plane spacings and strains in compressed solids; the methods are quite varied: white synchrotron radiation with Laue diffraction from single crystals have been used to record the full strain tensor [12] although the shear strains are not always determined [13], synchrotron white radiation with a collimated X-ray beam, an energy dispersive detector and Laue diffraction [14] are reported to establish the full strain tensor, synchrotron radiation with Debye-Scherrer rings from polycrystalline materials [15-18] may also establish the strain, flash X-ray sources with image plates and Debye-Scherrer rings from polycrystalline materials [19-23] and X-ray free electron laser sources with single crystals and/or polycrystalline materials [24] identify diffracted rays for strains and phases in compressed materials. The measured plane spacings from these experiments in-turn determines the lattice strains and the strain tensor's components.

Conventional, commercial diffractometers fitted with Eulerian cradles for adjusting the planar $\psi$ angle i.e., a rotation about an axis lying in the plane of the sample's surface, is able to measure two dimensional strain tensors on flat surfaces [25]; two dimensional detectors with $\theta-\theta$ or $\theta-2\theta$ diffractometers, also on flat samples, have software [26] for finding two dimensional strain tensor components from diffracted Debye-Scherrer ellipses.

The stresses from planar, longitudinal waves [8] in solids are assembled through a mechanics identity in equation (1):

$$\sigma_{11} = -p + 2\tau \qquad (1)$$

$\sigma_{11}$ is the principal stress in the direction of wave propagation; $p$ is the pressure defined from the trace of the stress tensor; $\tau$ is an octahedral shear stress as given in equation (2).

$$p = -\frac{(\sigma_{11} + \sigma_{22} + \sigma_{33})}{3}; \quad \tau = \frac{(\sigma_{11} - \sigma_{22})}{3}. \qquad (2)$$

Substitution of equation (2) into equation (1) plus $\sigma_{22} = \sigma_{33}$ i.e., assuming the material is isotropic with $\sigma_{22}$ and $\sigma_{33}$ as principal stresses, establishes expression (1) as a mathematical identity that organizes the stress tensor components into pressure and shear terms for planar stress waves in longitudinal compression. Equation (1) contains no physics but is an organizational aid for relating the stress in the wave direction, $\sigma_{11}$, to $p$ and $\tau$.

The most widely used expression in (1) is $\tau = 0$ and the solid material is "fluid-like." A pressure description allows for an equation of state in the solid [27-29] and relations between specific volume, temperature, pressure, and entropy. Equations of state are complete descriptions of the material's behavior excluding shear stresses. However, at significantly lower stresses there are systems with the wave being linear elastic and Poisson's effect describing the material's elastic stress response. The material's behavior is ideally somewhere between these two extremes.

Ablative, planar, dynamic, longitudinal waves [3, 4, 6, 19, 30-37] that are ramp-compressed, shocked, or doubly shocked, assume a stress state of hydrostatic compression. The solid material is "fluid-like" as the material is thought to only support small shear stresses. This restriction allows Rankine-Hugoniot equations [38-40] of a liquid to be applied to velocity interferometer systems for any reflector, VISAR, measurements [41, 42] and relate the solid's velocities to unidirectional stresses in the propagation direction; additionally, if the stresses are assumed hydrostatic then the Rankine-Hugoniot equations from fluids fully describe the pressure in the solid. Equations of state as noted earlier [43-49] can further exploit and characterize the "fluid-like" solid material. Thus, the measured compressive stresses in the direction of wave propagation when considered as a pressure implies, through equation (1) that the shear stresses for plasticity or phase changes are very modest, so pressure prevails. Experimental verification of the pressure supposition however is difficult to obtain. It requires several independent measurements of strain and isotropy assumptions.

Strain is imposed on material in a longitudinal compression wave [9, 10] as a uniaxial strain. Figure 1 shows the material in ambient conditions and the same material when pressurized or when deformed by a uniaxial strain. The deformation is assumed to be in the wave propagation direction; the strain perpendicular to the wave propagation direction is considered zero. A planar wave should have no deformation in the perpendicular direction as longitudinal elastic waves are constructed with uniaxial strains in the wave direction. The wave front as recorded in VISAR is planar. Figure 1 is a schematic of the two models with strain from pressure or a longitudinal strain compressing the lattice from ambient. Maintaining a planar wave front in the strained system is related to the deformation of the material.

Bragg's law construction in 3-dimensions for a collimated, monochromatic, X-ray beam acting on a reciprocal space lattice is in the next section; the Polanyi sphere in the reciprocal lattice is transformed into an ellipsoid of revolution by strain. Ewald's sphere in 3-dimensions intersects this ellipsoid on a curved surface. The intersection points of the two surfaces establish the diffraction cone which includes the point of the X-ray source. Knowledge of the crystal structure in the material and a second diffraction plane are used here to find two points on the ellipsoid.

These two planes, when analyzed, establish the strain tensor of the material. The strain imposed on the reciprocal lattice by the longitudinal wave is found from measurement of the diffraction conditions. Finally, we combine all our measurements of pressure, strain and material crystal structure to tabulate the mechanical strain response to applied stresses and compare the data to known models of strained solids in the conclusions section.

## *Methods*

*Reciprocal Space and Diffraction Descriptions of Uniaxial Compressed Material*

Figure 2 is a schematic of the laser ablation compression of materials used to explore material properties with details noted in other publications [20-22]. The PXRDIP camera uses imaging plates to record the diffracted X-rays. The X-ray source is external to the camera. The diffraction lines seen on the image plates in figure 2 are interpreted through Debye-Scherrer rings and detect atomic structures during nanosecond compression. These rings frequently contain differences from the material's texture but not strain differences in the lattice as noted earlier [12-16, 18, 50]. Material density differences from two different $(hkl)$ diffraction planes in the same material as noted in [20] establishes that hydrostatic strain alone is incapable of describing the material's strain tensor.

Debye-Scherrer diffraction cones from laser compression in what follows will be constructed from a strain-distorted reciprocal lattice, i.e. an ellipsoidal, Polanyi's surface which is intersected by Ewald's sphere. The compressed material expands the reciprocal lattice as it intersects the sphere formed by the vectors of incidents and diffraction as seen in figure 3. Polanyi's sphere with concentric spheres for each atomic plane is generated from a fully rotated polycrystalline reciprocal lattice representing $(hkl)$ planes. Uniaxial compressive strains in the direction of wave propagation, however, expands this reciprocal lattice in the wave direction generating an ellipsoid of revolution from the original spherical Polanyi shape. As seen in figure 3, the Bragg angle, $\theta$, changes between the unstrained and strained diffraction constructions. The diffraction in a full 3-dimensional space is found with the vector of diffraction, $\vec{k}$, touching any point on the now ellipsoidal Polanyi reciprocal space structure. $\vec{k}_0$ is the vector of incidence. The diffraction geometry is not Bragg-Brentano but rather Bragg-Laue as the diffracted ray typically passes through the specimen. Also, the figure shows that the $x^*$ axis in reciprocal space corresponds to a major compressive strain direction while the $y^*$ and $z^*$ axes are assumed minor strained directions. The $y^*$ axis is chosen to be in the plane that includes the point X-ray source and the $x^*$ axis of the compression wave. $y^*$ is not necessarily aligned with the physical vertical axis on the imaging camera. The following diffraction analysis is for the full 3-dimensional example as obtained in PXRDIP or TARDIS [51, 52] cameras. The recently designed flash or pulsed [53, 54] TRXRD cameras may be [55] also described by this diffraction analysis. These newer cameras may use detectors located near a left side port on the camera seen in figure 2.

*Selection of the Strained X-Ray Diffraction Plane*

Bragg's law in reciprocal space [56] is satisfied when the vector of incidence, $\vec{k}_0$ plus the reciprocal lattice vector, $\vec{r}\,*$ for a particular plane equals the vector of diffraction, $\vec{k}$. Consider the plane formed by the axis of the longitudinal wave passing through the centroid of the sample and containing the X-ray source. Let this plane be the x-y plane in real space. The wave vector of incidence, $\vec{k}_0$, is in this plane. Bragg's law is:

$$\vec{k}_0 + \vec{r}\,* = \vec{k} \qquad (3)$$

$\vec{k}_0$, terminates on the reciprocal space origin. The reciprocal lattice vector of a specific plane is $\vec{r}\,*$. Diffraction occurs when the sum of the vector of incidence plus the reciprocal lattice vector equals the wave vector of diffraction, $\vec{k}$. The representation seen in equation (3) is shown in figure 4a when the ray's path lengths differences from neighboring planes constructively add. The physical arrangement is shown in figure 4a with the incident wave vector striking the diffracting plane. Bragg's diffraction condition in equation (3) is shown in figure 4b in reciprocal space construction. Bragg's law selects the orientation of the plane that will diffract. In this case, the geometry on the right shows the algebraic form of Bragg's law seen in equation (4). From figure 4b we have:

$$k_0 \; Sin(\theta) = \frac{r\,*}{2} \quad \text{or} \quad \frac{Sin(\theta)}{\lambda} = \frac{1}{2d} \qquad (4)$$

Thus, Bragg's law in the form $\lambda = 2d \; Sin(\theta)$ is contained in equations (3) and (4). The reciprocal space representation in figure 4b is a geometric construction of Bragg's law. If $\vec{r}\,*$ were to lie in the x*-y* plane, then the diffracted X-ray represented by $\vec{k}$ is also in this plane.

*Longitudinal Wave Strain*

Planar, longitudinal wave propagation has a principal uniaxial strain aligned with the direction of wave propagation. Let $\varepsilon_{11}$ be the true strain in the direction of wave propagation. $\varepsilon_{11}$ represents the change in the interatomic plane spacing in the longitudinal wave direction. Polanyi's sphere when strained is changed into an ellipsoid of revolution by this strain:

$$\varepsilon_{11} = \int_{d_0}^{d_1} \frac{dx}{x} = \ell n \frac{d_1}{d_0} \qquad (5)$$

The interatomic plane spacing is initially $d_0$, with compression, however, it is compressed to $d_1$ in the wave direction. The engineering strain from equation (5) is $\Delta d/d_0$ with $\Delta d$ the change in the plane spacing from compression. The true and engineering strain are related by

$$\varepsilon_{11} = \ell n \frac{d_1}{d_0} = \ell n \frac{(d_0 + \Delta d)}{d_0} = \ell n(1 + \frac{\Delta d}{d_0}) \approx \frac{\Delta d}{d_0} - (\frac{\Delta d}{d_0})^2 + \; ..... \qquad (6)$$

The engineering strain is the first term in a series expansion of the true strain. Figure 4a is a schematic of the strain generated by the displacements seen in figure 1 on the right. The material perpendicular to the wave propagation direction may also be strained. A more inclusive model as developed below investigates both pressure and uniaxial strain. Pressure is represented in the middle in figure 1. The pressure is modeled to have an atomic spacing of $d_2$. It will account for any in-plane strain. The pressure model in figure 1 is uniformly strained. The strain in the plane of the wave is:

$$\varepsilon_{22} = \int_{d_0}^{d_2} \frac{dx}{x} = \ln \frac{d_2}{d_0} \qquad (7)$$

The strain tensor from figure 1 that will include both pressure and uniaxial strain is:

$$\begin{pmatrix} \varepsilon_{11} & 0 & 0 \\ 0 & \varepsilon_{22} & 0 \\ 0 & 0 & \varepsilon_{22} \end{pmatrix} \qquad (8)$$

Strains from both the pressure and the uniaxial strain models in figure 1 are both included in the strain tensor in equation (8). Thus, experimental measurements if supporting $\varepsilon_{22} = 0$ would indicate only uniaxial strain; if experimental observations have $\varepsilon_{11} = \varepsilon_{22}$ then pressure models are favored. Strain measurements can aid in understanding the material's deformation. The atomic plane spacings are between $d_1$ and $d_2$ depending on the strain's directions in equations (5) to (8) and the plane's orientation to the compression axis.

As the material is compressed, the reciprocal space is expanded in the direction of the strain. The "d" spacing in equation (9) depends on the orientation of the coordinates of $\vec{r}*$.

$$\frac{1}{d} = |\vec{r}*| = \sqrt{(x*)^2 + (y*)^2 + (z*)^2} \qquad (9)$$

The sphere from the vector of incidence and the vector of diffraction intersects Polanyi's ellipsoid as seen in figure 3. The directions perpendicular to the propagating wave are strained with $\varepsilon_{22}$ so Polanyi's original sphere is transformed to an ellipsoid of revolution. This lattice is generated from the original isotropic polycrystalline sphere expanded by the compressive strain applied to the reciprocal lattice in the direction of the shock wave and perpendicular to it as described in equations (5) to (8). It is well known in solid mechanics that a sphere when strained deforms into an ellipsoid.

The equations of intersection of Ewald's sphere and the ellipsoid of revolution are found by construction of the sphere of diffraction centered at the tail of $\vec{k}_0$. This vector is along the ray's path from the ionized X-ray source to the centroid of the target. $\vec{k}_0$ makes an angle $\alpha$ with the $x*$-axis and lies in the $x*$-$y*$ plane. The diffraction in 3-dimensions is from the solution of the intersection of the sphere of diffraction, formed using the vector of incidence:

$$\vec{k}_0 = \frac{Cos\alpha}{\lambda}\hat{i} - \frac{Sin\alpha}{\lambda}\hat{j}. \qquad (10)$$

The Ewald sphere formed by $\vec{k}$ has the vector's tail located at the tail of $\vec{k}_0$.

$$(\lambda x^* + Cos\alpha)^2 + (\lambda y^* - Sin\alpha)^2 + (\lambda z^*)^2 = 1. \qquad (11)$$

Equation (11) represents Ewald's sphere, in reciprocal space. It is offset from the origin as seen in (11).

The reciprocal space ellipsoid represents the strain-compressed lattice centered at the origin with a major axis of $1/d_1$ and a minor-axes of $1/d_2$. The surface is an ellipsoid of revolution.

$$(d_1 x^*)^2 + (d_2 y^*)^2 + (d_2 z^*)^2 = 1. \qquad (12)$$

The points $X^*$, $Y^*$ and $Z^*$ are the intersection points from both surfaces. In general, the 3-dimensional solution is a closed curve that defines a conic section with the X-ray source. The points on the surface of the cone are distances of $1/\lambda$ from the tip of the cone. Debye-Scherrer diffraction rings are circles in unstrained materials. In strained materials, the diffraction rings are distorted as the $1/d$ distance changes with orientation so Bragg's angle is strain dependent. The intersection diffraction structure contains major and minor axes. It remains to be established if the cross-section of the conic-section is an ellipse. An ellipsoid of revolution intersected by a flat plane does form an ellipse [57]; the Ewald sphere is curved so the intersection with the reciprocal space ellipsoid is not a flat plane. The intersection structure has both major and minor axes, but the structure's curve is unlikely to be elliptical. The strain distorted reciprocal space, however, is an ellipsoid of revolution.

The intersection points $X^*$, $Y^*$ and $Z^*$ form a 3-dimensional reciprocal space curve. The vectors of diffraction $\vec{k}$, point from the tail of $\vec{k}_0$ in $\vec{r}^*$ space to the curve defined by the points in the solutions from (11) and (12).

The angle between the vector of incidence and the vector of diffraction is $2\theta$:

$$\vec{k}_0 \cdot \vec{k} = \frac{Cos(2\theta)}{\lambda^2} \qquad (13)$$

$2\theta$ is ½ of the total angle of the distorted Debye-Scherrer cone. Two examples below provide guidance to what is expected from experimental data when recorded on image plates or 2-D detectors. These instruments measure the "d" spacings from Bragg's law by direct measurement of $2\theta$ for a specific diffracted plane's "d" spacings with a known wavelength.

Solutions to equations (11) and (12) are easily understood in the limit when $\lambda \to 0$. In this approximation the Ewald sphere of diffraction is a flat plane containing the reciprocal lattice origin. The intersection structures are now ellipses. The first diffraction solution with $\alpha = 0$ is a plane that is perpendicular to the incident X-ray, containing the origin of the reciprocal lattice. It is a circle in reciprocal space with a radius of $1/d_2$. The Debye-Scherrer cone is thus only from in-plane strained material. This very short wavelength approximation can also be applied to $\alpha = 90°$; the diffraction solution in reciprocal space is a full ellipse with both $1/d_1$ and $1/d_2$ on the major and minor axes seen in Figure 3. Debye-Scherrer's cone in this case contains the eccentricity imposed by the ellipsoid. Thus, in the limit $\lambda \to 0$ the compressive strains and the interatomic plane spacing due to longitudinal waves are both observable. If the shear strain is zero, then $1/d_1$ and $1/d_2$ will have the same value and the material would be in hydrostatic strain with all normal strains equal.

## *Results*

*Diffraction for Measuring the Strain Tensor*

The strain state is imbedded in the ellipsoid of revolution. The strains in equations (5) to (8) are from the spacings of $d_1$, $d_2$ and $d_0$. Two independent measurements of two points on the ellipsoid determines $d_1$ and $d_2$. Note that all *(hkl)* planes have a range of spacings. Figure 5a shows an annulus of compression in reciprocal space with the inner radius $1/d_0$ and the outer radius of $1/d_1$. The trace of the ellipsoid shown represents the reciprocal plane's spacings from the imposed uniaxial strain and the assumed zero in-plane strain. The ellipse is tangent to both circles and describes the *(hkl)* plane's spacing relative to the compression axis and distance from the origin.

Figure 5b shows two points on the ellipsoid. The two points are $p_1(X_1^*, Y_1^*)$ and $p_2(X_2^*, Y_2^*)$. Both points are diffraction points and satisfy Bragg's law so,

$$(X_1^*)^2 d_1^2 + (Y_1^*)^2 d_2^2 = 1 \quad \text{and} \quad (X_2^*)^2 d_1^2 + (Y_2^*)^2 d_2^2 = 1 \qquad (14)$$

Equation (14) has two unknowns, $d_1^2$ and $d_2^2$ with two equations. They are linear equations thus, knowledge of the points $p_1$ and $p_2$ determine the unknowns with the aid of linear algebra. If the two points include a component with $Z^*$ then the equations will still represent a linear algebra problem.

Figure 5c establishes the selected geometry for point $p_1$ with Bragg angle $\theta_1$, from the arrangement of the X-ray source, Bragg's law, the diffraction plane spacing and the ellipsoid. $X_1^*$ and $Y_1^*$ as obtained using the geometry in figure 5c. Note that $X_1^*$ and $Y_1^*$ are the sine and cosine of the angle $\beta$ respectively with $1/d$ so:

$$X_1^* = -\frac{2 Sin(\theta_1) Sin(\theta_1 - \alpha)}{\lambda} \quad \text{and} \quad Y_1^* = \frac{2 Sin(\theta_1) Cos(\theta_1 - \alpha)}{\lambda}. \qquad (15)$$

Bragg's law was included in equation (15). Substitution of equation (15) into equation (14) on the left gives one equation with two unknowns.

The second point $p_2$ can be determined from a second diffraction plane say $(h'k'l')$. Figure 5d establishes in reciprocal space the second plane relative to the $(hkl)$ plane.

The diffraction planes $(h'k'l')$ and $(hkl)$ are not parallel but the two planes on the minor and the major axes are parallel; the fully compressed planes at the major points on the two ellipsoids have the same strain. The uncompressed plane spacings before compression can be related to the material's crystallography either analytically or experimentally using measurements, see the International Center Diffraction Data [58] or other crystallographic databases for plane spacings.

The $(h'k'l')$ and the $(hkl)$ plane spacings before compression are related by

$$d_{0'} = c_R d_0. \qquad (16)$$

Crystallography relates the primed plane spacing to the unprimed spacing as $d_{0'} = c_R d_0$. An example with the cubic planes of (111) and (200) is $c_R = \sqrt{3}/2$. $c_R$ is the ratio of the $(h'k'l')$ plane's unstrained spacing to the $(hkl)$ plane's unstrained spacing.

The lattice strains in equations (5) and (8) are aligned with the $x^*$ and $y^*$ axes respectively. The $(h'k'l')$ plane's spacing to the $(hkl)$ plane's spacing from the strain in equation (5) with equation (16) gives the first expression below on the left with equation (8) and (16) to give the term on the right:

$$d_3 = c_R d_1 \text{ and } d_4 = c_R d_2 \qquad (17)$$

The $(h'k'l')$ ellipse from equation (14) on the right thus becomes,

$$(X_2^*)^2 d_3^2 + (Y_2^*)^2 d_4^2 = 1 \text{ or } (X_2^* c_R)^2 d_1^2 + (Y_2^* c_R)^2 d_2^2 = 1 \qquad (18)$$

Thus, diffraction from two planes, equation (14) on the left with equation (15) and equation (18) on the right determines the solution of the compression axis ratio using linear algebra. Bragg's geometry from equation (14) and the second Bragg angle, $\theta_2$, of the $(h'k'l')$ plane yields:

$$\left(\frac{d_1}{d_2}\right)^2 = \frac{(c_R \sin\theta_2 \cos(\theta_2 - \alpha))^2 - (\sin\theta_1 \cos(\theta_1 - \alpha))^2}{(\sin\theta_1 \sin(\theta_1 - \alpha))^2 - (c_R \sin\theta_2 \sin(\theta_2 - \alpha))^2} \qquad (19)$$

Note that $\alpha$, the two Bragg angles, $\theta_1$ and $\theta_2$ plus the unstrained crystallography relation, $c_R$, determines the compression axis ratio of the two perpendicular directions, $d_1/d_2$. If the material is in hydrostatic pressure, as discussed next, then the ratio of the plane spacings from experimental measurements in all directions in equation (19) should have $d_2 = d_1$ and equation (19) would then represent circles not ellipses for both the $(hkl)$ and $(h'k'l')$ diffraction planes.

*X-Ray Diffraction from Material with Hydrostatic Pressure*

The wave propagation model with a stress state of hydrostatic pressure has $\tau$ zero in equation (1). The three principal stresses are all equal and the stress in the $x^*$ direction is equal to the pressure from equation (1). X-ray diffraction measures plane spacings which are used to obtain interatomic plane spacings and strains which for hydrostatic pressures are all taken to be equal. The measurement by diffraction for hydrostatic strains will have $d_1/d_0$ the same in all directions. The material is compressed with equal displacements for all three directions. The Debye-Scherrer diffraction cones will show no eccentricity with hydrostatic pressure. This implies that the in-plane strain of the wave front which is perpendicular to the propagation direction is no longer a planar strain as shown in figure 1. Discussion of measured strains are embedded in the descriptions in Table I and II below. The two or more diffraction planes discussed with a compressed materials list from [19-22, 59] have been reported in the literature.

*Materials Previously Studied for Phase Transformations*

Diffraction patterns from wide selections of materials and laboratories have been made using laser ablation. See work from: Laboratory for Laser Energetics, Lawrence Livermore National Laboratory, Rutherford Appleton Laboratory, European X-Ray Free-Electron-Laser laboratory, and others that have reported material crystal structures. Measurements at LLE are recorded with X-ray diffraction patterns on image plates from diffraction lines and simultaneous timed VISAR measurements of pressure as seen in figure 2. The quantitative measurements were obtained by integration around the $\varphi$ or azimuthal dependence of the Debye-Scherrer ring, by assuming circles for obtaining the intensity versus $2\theta$ which are reported as 'lineouts' or intensities versus $2\theta$. Plane spacings and crystal structures are determined from lineout evaluations. Equation (18) however is based on X-ray intensity measured at two independent points both on the reciprocal lattice with the ellipsoid formed by lattice compression. In what follows below, the Bragg diffracted peaks reported from selected LLE's lineouts measurements for two planes within the same crystalline phase are used. Established crystal structures provide the $c_R$ values used to relate the second plane spacing to the first planes spacing. The four materials investigated are individually listed below and in Tables I and II.

Aluminum [20-22, 60] was shocked-compressed to increasing pressures in separate experiments while simultaneously recording the fcc, hcp and bcc crystalline phases, pressures, and individual Bragg peaks. VISAR derived pressures were calculated using Rankine-Hugoniot relations for a fluid in equation (1) with $\tau = 0$. The wavelengths of the helium like X-ray sources were from foils with $\lambda = 1.48 \text{\AA}$ for Cu and $\lambda = 1.21 \text{\AA}$ for Ge. The selected Bragg diffracted beams, $2\theta$ were imaged at 45.49° and 51.74° for the hcp phase with the (100) and (101) diffraction planes. The bcc phase has only a single recorded peak from the (110) plane at $2\theta$ so it is not used.

Nickel is known to remain face-centered-cubic up to several TPa of pressure. The fcc's (111) plane has $2\theta$ Bragg angles of 63.3° and 69.9°, and was selected at 250 GPa and 402 GPa respectively [59]. Simultaneously, the fcc's (200) plane had $2\theta$ Bragg angles of 74.9° and 77.6° at 250 GPa and 402 GPa. Both diffraction planes were recorded with a helium-like Fe foil from a backlighter, X-ray with $\lambda = 1.855 \text{\AA}$ at an angle of $\alpha = 22.3°$. $c_R = 0.8660$ between the (111) and (200) planes as noted in equation (16). These values with equations (5) and (18) yield a uniaxial strain $\varepsilon_{11}$ of -0.825 from a compression ratio of 0.438 and an engineering shear strain angle of 39.5°.

Sodium is a very compressible metal with a large thermal expansion coefficient. The thermal expansion coefficient couples the isentropic temperature to the pressure through the thermoelastic [61-64] coefficient. The outer Na shell contains a single valance electron in the 1s orbital. This free valance electron at low pressure aides' good conductivity. However, at high pressure Na forms electride structures [19, 65] when pressures approach 100's GPa. The electride phase electrons are resident in the interstices of the layered atomic planes. The measured hP4 phase can reduce the c/a ratio in the cell which increases the density as reported by [19]. Density increases of 7 times ambient have been recorded at high pressures in Na; strained materials have heterogeneous densities, so compression ratios are important to sort out the density of the phase. Four diffraction peaks in the hP4 phase have been recorded: (010), (011), (012), and (110) at $2\theta$ values of 36.1°, 43.3°, 60.8°, and 65.4° respectively. The X-ray wavelength is $\lambda = 1.47 \text{Å}$. The pressure was measured at 409 GPa and $\alpha = 22.5°$ was reported.

Invar [66] is an alloy of nickel-iron with 36 weight % Ni in a ferrous alloy. It was first discovered by C-E. Guillaume [67] for which he received the 1920 Nobel Prize in physics. The diffraction properties are summarized in Table I. In the cubic phase (111) and the (200) planes diffract at $2\theta$ Bragg angles of 63.8° and 73.2° at pressure of 367 GPa. $\alpha = 22.3°$.

Table I X-ray diffraction data for selected materials from longitudinal compression experiments.

| Material | Crystal Structure | Atomic Plane | Bragg's angle, $2\theta$ | Compression axis to Source angle, $\alpha$ | Equation (18) $d_1/d_2$ |
|---|---|---|---|---|---|
| Aluminum | hcp | (100) | 45.49° | 22.3° | NA |
| c/a=1.65 | hcp | (002) | 48.00 | 22.3° | 0.576 |
| | hcp | (101) | 51.75 | 22.3° | 0.713 |
| Nickel | fcc | (111) | 65.9° | 22.3° | NA |
| | fcc | (200) | 77.6° | 22.3° | 1.048 |
| Sodium | hP4 | (010) | 36.1° | 22.5° | NA |
| a, b = 2.75 Å | hP4 | (011) | 43.3° | 22.5° | 1.32 |
| c/a=1.35 | hP4 | (012) | 60.8° | 22.5° | 0.619 |
| | hP4 | (110) | 65.4° | 22.5° | 0.694 |
| Invar | | | 63.8° | 22.3° | NA |
| | | | 73.2° | 22.3° | 1.577 |

Table II Mechanical properties in longitudinal wave compression experiments.

| Material | Pressure, GPa | Compression Axis Ratio, $d_1/d_2$ | Uniaxial Strain, $\varepsilon_{11}$, equation (5) | Engineering Shear Angle, |
|---|---|---|---|---|
| Aluminum, hcp | 291 | 0.65 | -0.11 | 6.3° |
| Nickel, fcc | 402 | 1.048 | -0.73 | 36° |
| Sodium, hP4 | 409 | 0.878 | -0.46 | 25° |
| Invar, fcc | 367 | 1.577 | -0.32 | 18° |

Tables I and II are for selected data from three different materials with up to three phases as the pressure increases. The compression axial ratios with the aid of equation (5) gives the uniaxial strain, $\varepsilon_{11}$. The uniaxial strain is directly related to the shear angle for the strain tensor when rotated 45° from the principal axes. The angles reported in Table II are engineering shear strain angles. Materials that are pressurized or stressed more than the bulk modulus will have non-linear elastic responses.

Strain is a measurable quantity that describes atomic displacement positions in materials. The two compression strains are seen in equation (8); they describe material deformation. The major purpose of this manuscript is to clarify differences between the two major models as shown in figure 1: a pressure description and/or a strain description. Although Table I and II aid in that objective, the full data set for the high purity aluminum material will now be included as is discussed below.

The compressive strains for an aluminum data set are from the diffraction data in [20]. $\varepsilon_{11}$ and $\varepsilon_{22}$ are from the $d_0, d_1, d_2$ planes and equations (5) and (7). The results for the strain $\varepsilon_{22}$ are shown in figure 6a. Strain versus pressure is from two of the three aluminum phases while the results show significant scatter, it establishes a clear trend: increasing the pressure, decreases, the $\varepsilon_{22}$ strain. The value of $d_2/d_0$ is used in the strain of equation (7) in figure 6a and it shows significant scatter because it is from the difference between plane spacing. The experimental data used in equations (14), (15) and (18) came from a program that presumes diffraction is always from material in hydrostatic strain. The value of the strain, $\varepsilon_{11}$, in the wave direction is inconsistent with 6a measurements and is shown in 6b. This shows that $\varepsilon_{11} \neq \varepsilon_{22}$ so a state of hydrostatic strain is not consistent with the data measured, even with the data biased in that direction. The linear fit in figure 6a passes very close to the origin and has been made to pass through the origin. All the data used to describe strains were obtained assuming that planes' spacings are unaffected by the assumption of hydrostatic pressure used in the software package [68] HEXRDGUI; unfortunately, this adds a bias to the strains. $\varepsilon_{11} \neq \varepsilon_{22}$ however, $\varepsilon_{11}$ data versus pressure is very poorly described as seen in figure 6b. The reason is deeper than just the use of a biased software package for the measurements. When compared to figure 6a it is apparent that $\varepsilon_{11}$ is inconsistent with any models and our understanding of all material behavior. The material is surely in compression but the strains aside from the two points to the left of zero strain are mostly in expansion which violates the first law of thermodynamics, as the material is compressed, and deformation is in expansion. This is expanded below with Eulerian and Lagrangian coordinates.

Linear algebra was used to evaluate $d_1$ and $d_2$; these values below are given using the $X^*$ and $Y^*$ values in the algebraic solutions.

$$d_1^2 = \frac{(Y_2^* c_R)^2 - (Y_1^*)^2}{(X_1^* Y_2^* c_R)^2 - (Y_1^* X_2^* c_R)^2} \text{ and } d_2^2 = \frac{(X_1^*)^2 - (X_2^* c_R)^2}{(X_1^* Y_2^* c_R)^2 - (Y_1^* X_2^* c_R)^2} \quad (20)$$

$d_1$ is related to the $Y^*$ positions of the Bragg diffracted X-ray divided by $X^*, Y^*$ terms. $d_2$ is related to the $X^*$ positions of the Bragg diffracted X-ray divided by the same $X^*, Y^*$ terms. The left side of equation (20) shows that the difference between the $(h'k'l')$ planes' $Y_2^*$ position and $(hkl)$ planes' $Y_1^*$ position determines $d_1$. The $Y^*$ values as seen in Figure 3 are found from the positions of the diffraction plane in reciprocal space. The aluminum data set selects diffraction planes that are not tilted close to the compression axis. Equation (20) establishes that the denominator which is the same for $d_1$ and $d_2$ is valid in both expressions. The numerator ideally should have large variability in $Y_1^*$ and $Y_2^*$ values. The $\varepsilon_{11}$ strain is not well determined by the slight direction differences of the planes and the diffracted X-ray beam. The $d_2$ plane spacing seen on the right of equation (20) uses the $X^*$ values of the reciprocal space which is a better reflection of the compression due to the wave. Equation (20) shows that it is not the denominator that gives scatter to $\varepsilon_{11}$ but the variability of the $Y_1^*$ and $Y_2^*$ data. Longer wavelengths would increase the Bragg $2\theta$ angle and thus make the diffraction planes more perpendicular to the compression axis.

*Eulerian versus Lagrangian Coordinates on the $X^*$ Axis*

The compressed specimen moves in a direction parallel to the *x* axis. The image plates are stationary in space as they are part of the PXRDIP or TARDIS cameras; the calibration diffraction lines are also stationary. In photography, an image is smeared unless the camera is panning the moving object. To record a moving object the correction from a stationary or Eulerian coordinate to a moving or Lagrangian coordinate is used in many technologies including: cinematography, thermal imaging [69] and optical radiometry. Application of the coordinate transformations to the expressions in (20) changes $\beta$ and is thus dependent on the relative material velocity, $V_c$, as measured by VISAR. The program [68] HEXRDGUI partly corrects for specimen motion in $2\theta$ but not in $\alpha$. The sample moves about 1 mm during total compression, while the X-ray flash pulse is shorter, with motion of the specimen less than 500 $\mu m$ during the X-ray image formation. See Figure 2. Absolute angle changes in equations (14), (15) and (18) due to the displacement of the sample at the time of exposure and are important and cited as the reason why measured differences in plane spacings fail to establish $\varepsilon_{11}$ strains. In the reciprocal space, $X^*$ changes so equation (20) is altered. The strain, $\varepsilon_{11}$ is modified by sample motion. Figure 6b is the difference between the $X^*$-component of reciprocal plane spacings used to obtain the strain $\varepsilon_{11}$. The VISAR motion is included in the recording in the program HEXRDGUI; the corrections to $\varepsilon_{11}$ await more precise measurements of the $X^*$-component from the diffracted plane to the reciprocal plane spacings.

*Strain Deformation*

Heterogeneous strains immediately leads [70, 71] to established deformation mechanism that relieve differences between strains. Dislocation glide [72, 73] is the most common mechanism for the fcc or hcp aluminum materials to deform irreversibly. Deformation with shear present in fcc structure [73] would typically be by dislocations established using Thompson's tetrahedron: the (111) glide plane has $\langle 1\bar{1}0|$, $\langle 0\bar{1}1|$, $|\bar{1}10\rangle$ total dislocations with partial dislocation of $|\bar{1}2\bar{1}\rangle$, $|\bar{1}\bar{1}2\rangle$, $\langle 2\bar{1}\bar{1}|$ and a $\delta$ intrinsic stacking [74] fault. The notation $\langle 1\bar{1}0|$ is used, instead of the usual $[1\bar{1}0]$ to indicate the sense of direction. If the deformation is in the $[111]$ direction then anticipate a Shockley partial dislocation of $\frac{1}{6}[211]$ would form with an intrinsic $\delta$ stacking fault. If the structure were instead hcp, then the deformation rules would follow [75] for the total and partial dislocations with intrinsic lattice faults.

The strain tensor trace is related to the specific volume change by

$$\frac{\Delta v}{v_0} = Tr(\varepsilon_{ij}) = \varepsilon_{11} + 2\varepsilon_{22} \qquad (23)$$

$Tr(\varepsilon_{ij})$ is the trace of the strain tensor and a value independent of the orientation of the strain tensor. $\Delta v$ is the change in the specific volume from, $v_0$, the initial state in compression.

## *Conclusions*

A new method for measuring strains with X-ray diffraction has been developed. The vector of incidence and a strain distorted reciprocal lattice were used to develop the principal strain tensor components. The strain tensor was embedded in a distorted reciprocal lattice. The diffracted X-rays were constructed using Bragg's diffraction geometry with the vector of incidence, an ellipsoid of revolution in reciprocal space, and the specific diffraction plane spacing. The reciprocal space geometry as measured by diffraction, leads to new expressions for residual strains.

Measured strains perpendicular to the stress wave shows that simple uniaxial strain does not describe laser ablated longitudinally compressed material at pressures close to a TPa. Strains in the direction of wave propagation unfortunately were not established due to motion of the material and the X-ray backlighter during compression. The developed method includes shear components that give inhomogeneous densities developed as strains although limited to the values between two circles seen in the annulus of compression in figure 5a. Thus, differences in density measurements for atomic planes as noted in [20] are explained based on inhomogeneous strains in compressed materials. The change in specific volume seen in equation (23) on the left for fcc Al is a strain of -0.5 at 238 GPa from "d" spacing data; so $\varepsilon_{11} = -0.35$ while $\varepsilon_{22} = -0.075$. Thus, significant shear strains exist in the compressed material at this pressure.

The software that was used for image evaluation and newer Python based software developed in [68] i.e., HEXRDGUI, assumes the Debye-Scherrer ellipses are circles. Furthermore, this software considers all diffraction planes to be homogeneous, at the same Bragg angle, and measurable throughout the entire sample. Bragg's law for a collimated, monochromatic beam selects only a single diffraction plane orientation (in 2-D, in 3D it is the Debye-Scherrer ellipse) relative to the strain tensor. This unique plane is at a specific angle to the compression axis, or the ellipse is at an angle to the strain -- a concept used throughout with the strain model developed.

Measuring 'two-points-on-the ellipsoid' was modified to measure strains from two diffraction planes, $(hkl)$ and $(h'k'l')$ in the same crystal structure. Ewald's sphere construction in two-dimensions intersects one point on each of the annuli of compression. Strain measurements from two planes with large differences in $2\theta$ values aid in improved data for superior strain resolution measurements. Interatomic plane spacings are dependent on the angle made to the strain axis. Especially when the $\varphi$ dependence of the Debye-Scherrer ring has been suppressed. Even with this knowledge $2\theta$ values above 90° are typically not available in the data used. Some $(\theta-\alpha)$ values are close to zero in the $Sin(\theta-\alpha)$ terms in equation (15). When they are small in the denominator, the numerator can be large.

The Bragg angles in this work have been measured at zero azimuthal angle and thus on the x*-y* plane in diffraction space. All the measured values are from 'lineout' measurements using the software noted above; it assumes Debye-Scherrer circles are without $\varphi$ or azimuthal angle dependence in $2\theta$. The strain constructions proposed are based on ellipses (not circles) as formed by Bragg diffraction from strained materials. An azimuthal dependence of atomic plane spacings is intrinsic in nearly all uniaxial strain concepts. The 'lineout' measurements based on circles smooth out elliptical structures and heterogeneous properties. The software thus biases strain measurements. Measurements from two independent planes remove some of the 'lineout' bias between the spacings with the equilibrium crystallographic atomic spacing values.

Strain measurements provide an independent method to establish if the concept of homogeneous pressure or heterogeneous strain is the more appropriate description of laser ablation for materials at TPa pressures. More importantly the strains describe the atomic positions and are very useful for understanding material behavior in laser compression. The overlap of atomic orbitals is a quantum driven strain-geometry problem. The longitudinal strains measurements made are considered heterogeneous values supporting a strain mixed with a homogeneous pressure model. For example, the strains measured in Na are less than the true compression strain. $\varepsilon_{11} = -0.65$ as measured is not 7-times density compression. Yet, Table II lists Na's strains as closer to -0.5, a very large strain but well below a strain of -2.0 from 7-times density compression.

The strain measurements made and seen in Table I and II plus Figure 6a have significant uncertainty: some materials show smaller strains while for other materials the strain values are larger. The difference between measurements accentuates the scatter. The extensive data sets from [20] with the present analysis emphasizes internal inconsistencies in $\varepsilon_{11}$. However, all $\varepsilon_{11}$ measured values are known to be tainted by pre-assuming the material is only in hydrostatic compression and is thus constructed from an assumed homogeneous deformation.

Strain and pressure or stress measurements of the compressive strains in material testing are generally relieved by plastically straining at the maximum shear stress angles to the uniaxial stress axis. Dislocations, twins, and most micro-faulting in irreversible deformation are totally mechanistically dominated by shear in all constitutive laws as is well-established in deformation literature. Shears alter the shape of the specimen. Shape changes are from measurable strains and a mix of planar strains with the uniaxial strains seen in figure 1. This occurs in strain measurements by Nature driving the material system towards a more homogeneous deformation state.

## *Acknowledgements*

SJB would like to thank Gilbert Collins, J. Ryan Rygg, Renato Perucchio, John C. Lambropoulos and Niaz Abdolrahim for discussions and encouragement. He would also like to thank the LLE students and staff for their weekly interactions. DNP would like to acknowledge that this material is partially based upon work supported by the Department of Energy National Nuclear Security Administration under Award Number DENA0003856, the University of Rochester, and the New York State Energy Research and Development Authority. The support of DOE does not constitute an endorsement by DOE of the views expressed in this article.

## *Author Contributions*

Both authors contributed equally.

## *Figures*

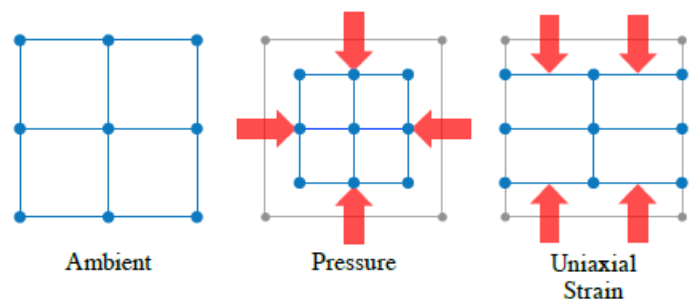

Figure 1. A schematic shows material in three conditions: ambient, pressurized, and with uniaxial strain. Ambient has no strains. Pressure is shown with homogeneous strains. Uniaxial strain displays heterogeneous displacements, so they are dependent on the compression axis.

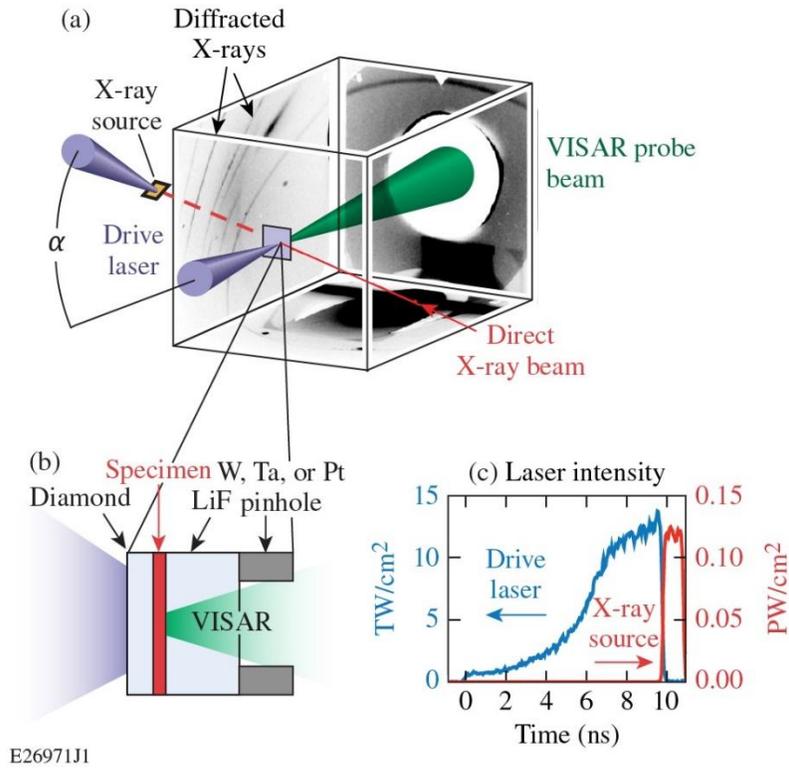

figure 2. (a) A 3-D schematic of the X-ray source utilizing a fully ionized K shell, the specimen, the VISAR probe beam, the X-ray image plates, and the drive laser are shown. Adapted with permission from the authors in [21, 51, 52]; note, the angle $\alpha$ is between the compression axis and the X-ray source. (b) The laser drive-beam illuminates a diamond crystal on the left which launches a shock into the specimen. The LiF window on the right is used by the VISAR beam to measure the velocity in the sample. (c) The drive laser and X-ray source timing are shown versus intensity.

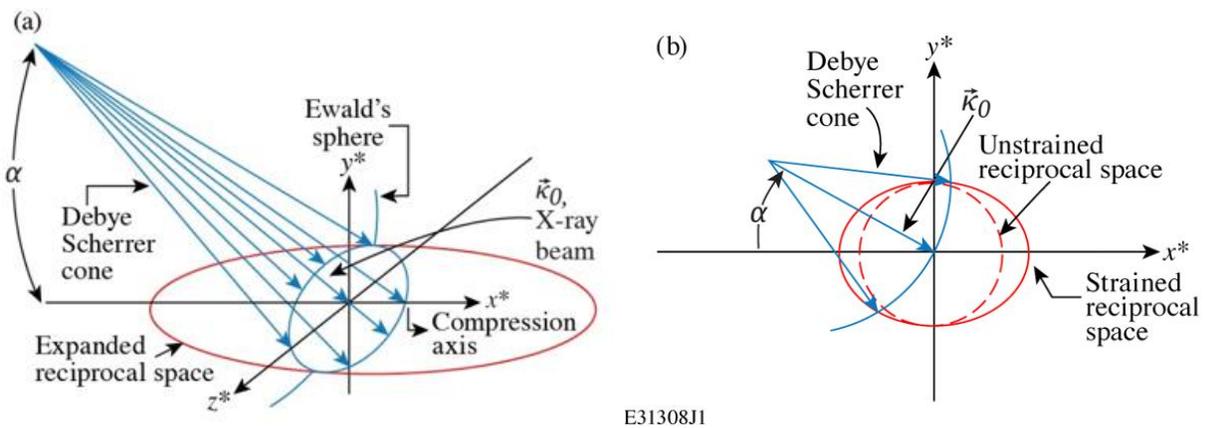

Figure 3. Reciprocal space construction of diffraction for a compressive strain in the longitudinal wave direction. $\vec{k}_0$ is the vector of incidence, $\vec{k}$ is the vector of diffraction, $x^*, y^*$ and $z^*$ are the reciprocal lattice axes. $2\theta$ is twice the Bragg angle between the vector of incidence and the vector of diffraction. (a) The compression axis is the $x*$ direction. The X-ray beam direction, the Ewald's sphere, the expanded reciprocal space and the Debye-Scherrer's cone construction are shown. (b) The $2\theta$ angle of the Debye-Scherrer cone depends on the difference between the unstrained, red-dashed sphere and strained solid-red ellipsoid of revolution. The major axis of the ellipsoid is aligned with the compression axis.

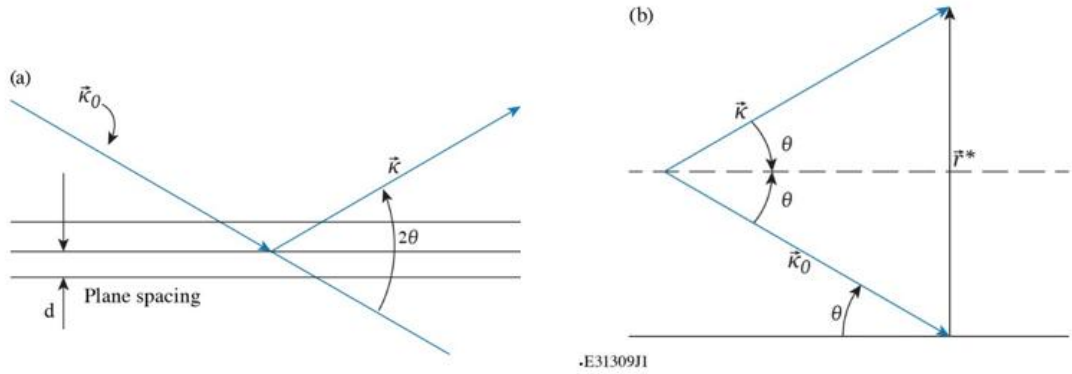

Figure 4. Braggs law in physical and reciprocal space. (a) The physical arrangement is shown with the atomic planes and the vectors of incidence and diffraction. The interatomic plane spacing "d" is shown. Bragg's angle is $\theta$ and the angle between the vector of incidence and diffraction is $2\theta$. (b) Reciprocal space construction is shown with Bragg diffraction. $\vec{r}*$ with a "1/d" spacing is the reciprocal lattice vector for the plane shown in (a).

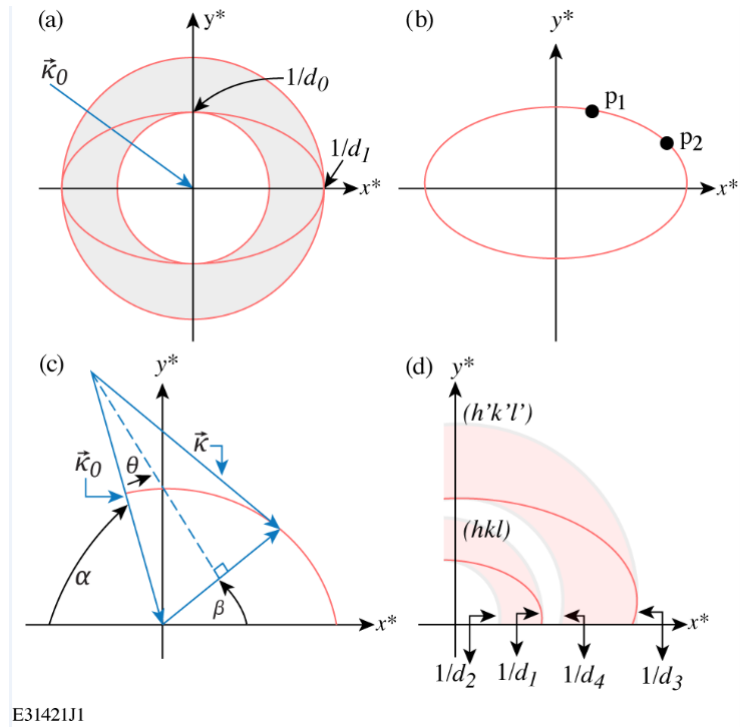

Figure 5. (a) The annulus of compression with an ellipse from the strained reciprocal lattice represents the uniaxial compressive strained lattice for an $(hkl)$ plane. (b) The equation of the ellipse can be determined by measuring two points on the ellipsoid with two independent diffraction conditions as shown. The points shown are in the x*, y* plane. The ellipse is represented by equation (12). (c) The angle $\beta$ is used to determine the values of $X_1^*$ and $Y_1^*$ with $1/d$. The geometry shows the compliment of $\theta$, $\alpha$, and $\beta$ add to $\pi$. (d) Again, only the 1/4 circle and ellipse are shown. The plane $(hkl)$ with spacings between $1/d_2$ and $1/d_1$ shows an ellipse between two circular arcs. The second plane $(h'k'l')$ with spacings compressed for $1/d_4$ to $1/d_3$ shows the ellipse between two circles. See the text for details.

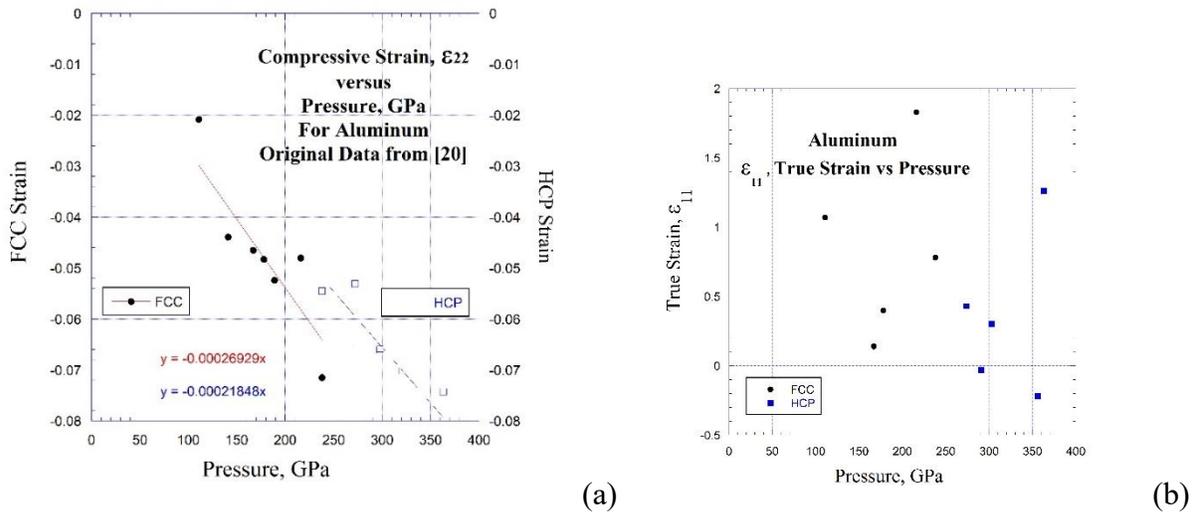

(a)            (b)

Figure 6. Pressure versus true strain. (a) $\varepsilon_{22}$, for aluminum obtained by measuring the difference in data from two diffraction planes to determine the ellipsoid of compression. The strain $\varepsilon_{22}$ versus pressure is represented by linear curves for fcc and hcp phases. The original X-ray plane spacings were measured in reference [20]. (b) The axial compressive strain, $\varepsilon_{11}$, versus pressure, $p$. The plot shows extensive scatter with inconsistent values. The strain $\varepsilon_{11}$ is measured from the differences in diffraction plane spacings in aluminum. See text for discussion.

5. Duffy, T.S. and R.F. Smith, *Ultra-high pressure dynamic compression of geological materials.* Front. Earth Sci., 2019. **7**: p. 1-20.
6. Kraus, R.G., et al., *Dynamic compression of copper to over 450 GPa: A high-pressure standard.* Phys. Rev. B, 2016. **93**(13): p. 134105-11.
7. Landau, L.D. and E.M. Lifshitz, *Theory of Elasticity*. 1959, London: Pergamon Press. pp. 98-103.
8. Burns, S.J., et al., *Planar, longitudinal, compressive waves in solids: Thermodynamics and uniaxial strain restrictions.* J. Appl. Phys., 2022. **131**: p. 215904-11.
9. Kolsky, H., *Stress waves in solids*. 1963: Dover Publications, Inc. pp. 4-40.
10. Graff, K.F., *Wave Motion in Elastic Solids*. 1975, London: Oxford University Press. pp. 273-310.
11. Dally, J.W. and W.F. Riley, *Experimental Stress Analysis*. 3rd ed. 1991, New York: McGraw-Hill, Inc. pp. 129-340.
12. Abboud, A., et al., *Single-shot full strain tensor determination with microbeam X-ray Laue diffraction and a two-dimensional energy-dispersive detector.* J. Appl. Cryst., 2017. **50**(3): p. 901-908.
13. Poshadel, A., P. Dawson, and G. Johnson, *Assessment of deviatoric lattice strain uncertainty for polychromatic X-ray microdiffraction experiments.* J. Synchrotron Radiat., 2012. **19**: p. 237-244.
14. Levine, L.E., C. Okoro, and R.Q. Xu, *Full elastic strain and stress tensor measurements from individual dislocation cells in copper through-Si vias.* Iucrj, 2015. **2**: p. 635-642.
15. Jiang, S., J. Zhang, and S. Yan, *Shear stress induced phase transitions of cubic Eu2O3 under non-hydrostatic pressures.* AIP Advances, 2023. **13**(2): p. 055308-8.
16. Zhang, H.J., et al., *Digital Image Correlation of 2D X-ray Powder Diffraction Data for Lattice Strain Evaluation.* Materials, 2018. **11**(3): p. 427-13.
17. Stavrou, E., et al., *Detonation-induced transformation of graphite to hexagonal diamond.* Phys. Rev. B, 2020. **102**(10): p. 104116-7.
18. Uzun, F., et al., *Extended caking method for strain analysis of polycrystalline diffraction Debye–Scherrer rings.* Crystals, 2024. **14**(8): p. 716-730.
19. Polsin, D.N., et al., *Structural complexity of ramp-compressed sodium to 480 GPa.* Nat. Commun., 2022. **13**: p. 2534-7.
20. Polsin, D.N., et al., *X-ray diffraction of ramp-compressed aluminum to 475GPa.* Phys. Plasmas, 2018. **25**(8): p. 082709-10.
21. Polsin, D.N., et al., *Measurement of body-centered-cubic aluminum at 475 GPa.* Phys. Rev. Lett., 2017. **119**(17): p. 175702-4.
22. Polsin, D.N., et al., *Errattum: Measurement of body-centered-cubic aluminum at 475 GPa (vol 119, 175702, 2017).* Phys. Rev.Lett., 2018. **120**(2): p. 029902(E).
23. Brown, N.P., et al., *DENNIS: a design and analysis tool for dynamic material x-ray diffraction experiments.* JINST, 2024. **19**(07): p. P07030-13.
24. Yang, H., et al., *Evidence of non-isentropic release from high residual temperatures in shocked metals measured with ultrafast x-ray diffraction.* J. Appl. Phys., 2024. **136**(5): p. 055901-9.
25. Noyan, I.C. and J.B. Cohen, *Residual Stress Measurement by Diffraction and Interpretation*. Materials Research and Engineering, ed. B. Ilschner and N.J. Grant. 1986, New York: Springer-Verlag. pp. 117-144.
20